\documentclass[11pt]{article}

\usepackage{graphicx}
\usepackage{amsmath}
\usepackage{amssymb}
\usepackage{dcolumn}
\usepackage{bm}
\usepackage[utf8]{inputenc}
\usepackage{authblk}

\newcommand{\be}{\begin{equation}}
\newcommand{\ee}{\end{equation}}
\newcommand{\ba}{\begin{eqnarray}}
\newcommand{\ea}{\end{eqnarray}}
\newcommand{\no}{\nonumber \\}
\newcommand{\gsim}{\mathrel{\hbox{\rlap{\lower.55ex \hbox {$\sim$}}
                   \kern-.3em \raise.4ex \hbox{$>$}}}}
\newcommand{\lsim}{\mathrel{\hbox{\rlap{\lower.55ex \hbox {$\sim$}}
                   \kern-.3em \raise.4ex \hbox{$<$}}}}

\def\vk{{\vec k}}

\def\roughly#1{\mathrel{\raise.3ex\hbox{$#1$\kern-.75em%
\lower1ex\hbox{$\sim$}}}}
\def\lsim{\roughly<}
\def\gsim{\roughly>}

\def\vx{{\vec x}}

\def\({\left(}
\def\){\right)}
\def\[{\left[}
\def\]{\right]}
\def\<{\langle}
\def\>{\rangle}
\def\pd{\partial}

\def\k{{\kappa}}
\def\l{{\lambda}}
\def\L{{\Lambda}}
\def\d{{\delta}}
\def\D{{\Delta}}
\def\o{{\omega}}
\def\O{{\Omega}}
\def\e{{\epsilon}}

\def\a{{\alpha}}

\def\g{{\gamma}}

\def\h{{\eta}}

\def\m{{\mu}}
\def\n{{\nu}}
\def\r{{\rho}}
\def\s{{\sigma}}

\def\th{{\theta}}
\def\ph{{\phi}}
\def\ps{{\psi}}

\def\x{{\xi}}

\newcommand{\wg}{{\wedge}}

\newcommand{\cN}{{\cal N}}
\newcommand{\MeV}{{\text{MeV}}}

\title{\bf Quark Mass Correction to Chiral Separation Effect and Pseudoscalar Condensate}

\author[2,3]{Er-dong Guo\thanks{guoerdong@itp.ac.cn}}
\author[1]{Shu Lin\thanks{linshu8@mail.sysu.edu.cn}}
\affil[1]{School of Physics and Astronomy, Sun Yat-Sen University, No 2 University Road, Zhuhai 519082, China}
\affil[2]{State Key Laboratory of Theoretical Physics, Institute of
Theoretical Physics, Chinese Academy of Sciences, Beijing 100190, China}
\affil[3]{Kavli Institute of Theoretical Physics China, Chinese Academy of Sciences, Beijing 100190, China}

\date{\today}

\begin{document}

\maketitle
%\vspace{0.1in}

\begin{abstract}
We derived an analytic structure of the quark mass correction to chiral separation effect (CSE) in small mass regime. We confirmed this structure by a D3/D7 holographic model study in a finite density, finite magnetic field background. The quark mass correction to CSE can be related to correlators of pseudo-scalar condensate, quark number density and quark condensate in static limit. We found scaling relations of these correlators with spatial momentum in the small momentum regime.
They characterize medium responses to electric field, inhomogeneous quark mass and chiral shift. Beyond the small momentum regime, we found existence of normalizable mode, which possibly leads to formation of spiral phase. The normalizable mode exists beyond a critical magnetic field, whose magnitude decreases with quark chemical potential.
\end{abstract}

\newpage

\section{Introduction and Summary}

The chiral magnetic effect (CME) \cite{Kharzeev:2007jp,Fukushima:2008xe,Kharzeev:2007tn} and chiral separation effect (CSE) \cite{Metlitski:2005pr,Son:2004tq} characterize the response of vector/axial current to the axial/vector chemical potential in external magnetic field. Both effects are manifestation of axial anomaly and are of phenomenological interest in heavy ion collision experiment. In particular, CME leads to charge separation and the interplay of CME and CSE gives rise to chiral magnetic wave (CMW) \cite{Kharzeev:2010gd}, which leads to charge dependent flow \cite{Burnier:2011bf}. There have been significant experimental efforts in search of CME \cite{Wang:2012qs,Adamczyk:2014mzf,Abelev:2012pa} and CMW \cite{Ke:2012qb,Shou:2014cua}, see \cite{Kharzeev:2015znc,Liao:2014ava,Huang:2015oca} and references therein. 

While CME and CSE share many similarities, they are known to differ in certain aspects. The chiral magnetic current is known to be independent from quark mass, temperature etc \cite{Fukushima:2008xe}. Correction may arise in dynamical cases, where axial chemical potential is not well defined and the dynamics of axial charge becomes important \cite{Iatrakis:2015fma,Wu:2016dam,Mueller:2016ven,Mace:2016shq}. The chiral separation current does not suffer from the issue of axial chemical potential, but it does receive correction from quark mass \cite{Metlitski:2005pr,Gorbar:2013upa,Chen:2013iga,Fang:2016uds,Kirilin:2013fqa}. In the static case, the correction to CSE can be derived in an ad-hoc way:
\begin{align}\label{anomaly}
\nabla\cdot {\bf j}_5=C \tilde{\bf E}\cdot \tilde{\bf B}+2M_q i\bar{\psi}\g^5\psi,
\end{align}
with $C=-\frac{N_ce^2Q^2}{2\pi^2}$. In the massless limit, we can write $\tilde E=-\nabla\m_q$, with $\m_q$ being the quark chemical potential. Since $\nabla\cdot\tilde{\bf B}=0$, we easily arrive at the celebrated CSE
\begin{align}\label{ad_hoc}
\nabla \cdot {\bf j}_5=-\nabla\cdot\({C\m_q \tilde{\bf B}}\)\Rightarrow {\bf j}_5=-C\m_q\tilde{\bf B}.
\end{align}
To obtain the mass correction to $j_5$, we first write the pseudoscalar operator $\s_5\equiv iM_q{\bar\psi}\g^5\psi$ as the response to quark chemical potential: $\s_5(x)=\int d^4yG_{\s5n}(x-y)\m_q(y)$. The structure of the Green's function $G_{\s5n}$ can be deduced from discrete symmetry: $\s_5$ is odd in both parity and time reversal, thus it should contain magnetic field $B$, which is odd in time reversal and spatial gradient $\nabla$, which is odd in parity. Therefore, to the lowest order in gradient, we have
\begin{align}\label{s5_corr}
\s_5=g(M_q^2,T,\m,\tilde B)\tilde{\bf B}\cdot\nabla\m_q.
\end{align}
Following the same steps as \eqref{ad_hoc}, we find correction to $j_5$,
\begin{align}\label{j5_corr}
{\bf j}_5=-C\m_q\tilde{\bf B}+2g(M_q^2,T,\m_q,\tilde B)\m_q\tilde{\bf B}.
\end{align}
The function $g$ is related to the Green's function as (in momentum space)
\begin{align}%\label{g_exp}
g=\frac{G_{\s5n}}{ik\tilde B}.
\end{align}
This relation will be confirmed analytically in model study. Note that $g$ vanishes when the quark mass $M_q$ vanishes. We can expand it in small $M_q$ regime:
\begin{align}\label{g_exp}
g=\#\frac{M_q^2}{T^2}+o(M_q^2)
\end{align}
assuming $\m_q\ll T,\; \tilde B\ll T^2$. The dimensionless prefactor $\#$ is to be determined by dynamics.
In fact, the analytic form of $g$ also constraints the response of $\s_5$ to quark mass $M_q$. Note that we have assumed a spatially inhomogeneous $\m_q$ and constant $M_q$. Instead, we can assume an inhomogeneous $M_q$ and constant $\m_q$. This should induce vev of $\s_5(x)=\int d^4yG_{\s5\s}(x-y)M_q(y)$. Consistency with \eqref{s5_corr} and \eqref{g_exp} indicates the following correction
\begin{align}\label{s5_dm}
\s_5=2\#\frac{M_q\m_q}{T^2}\tilde{\bf B}\cdot\nabla M_q+o(M_q^2),
\end{align}
which implies $G_{\s5\s}=2i\#M_qk\tilde B\m_q+o(M_q)$. We will provide clear numerical evidence for this correlator in model study. %These are the main results of the paper.

The correction to $j_5$ is more interesting in regime of large $\m_q$ and $\tilde B$. When $\tilde B=0$ and $\m_q$ large, different instabilities have been discussed in large $N_c$ field theory with spontaneous generation of chiral density wave \cite{Deryagin:1992rw,Shuster:1999tn}, current density \cite{Nakamura:2009tf,Ooguri:2010kt,Ooguri:2010xs} and quarkyonic spiral \cite{Kojo:2009ha,Kojo:2011cn,Kojo:2011cn,deBoer:2012ij} etc. At strong magnetic field, formation of chiral magnetic spiral \cite{Basar:2010zd,Kim:2010pu,Ammon:2016szz} is possible. Here we discuss a different type of instability characterized by pseudoscalar condensate. This instability is already identified in \cite{Kharzeev:2011rw} see also \cite{Brauner:2016pko} in low temperature confined phase. We extended the discussion and found it only exists within a window of magnetic field. Formation of this instability leads to spontaneous generation of chiral shift, first introduced in \cite{Gorbar:2011ya}, which induces further correction to $j_5$.

The paper is organized as follows: In Sec II, we give a brief review of the holographic model and the finite density and magnetic field background; Sec III contains a study of correlators among pseudoscalar condensate, quark condensate and quark number density in small momentum regime; Sec IV extends the study of correlators in arbitrary momentum regime and discussed the instability towards formation of spiral phase. We close the paper in Sec V with some outlooks.

\section{A brief review of the model}

\subsection{The finite density background}
We use the D3/D7 model to study the effect of finite quark mass. The
background consists of $N_c$ D3 branes and $N_f$ D7 branes. In the probe limit $N_f\ll N_c$, the background is simply given by black hole background sourced by D3 branes, with suppressed backreaction from D7 brane. The D3/D7 model is dual to ${\cal N}=4$ Super Yang-Mills (SYM) fields and ${\cal N}=2$ hypermultiplet fields, which transform in adjoint and fundamental
representations of $SU(N_c)$ gauge group respectively. By analogy with
QCD, we loosely refer to the ${\cal N}=4$ and ${\cal N}=2$ fields as
gluons and quarks respectively.
The black hole background of D3 branes is given by \cite{Mateos:2006nu}:
\begin{align}\label{d3_metric}
ds^2=-\frac{r_0^2}{2}\frac{f^2}{H}\r^2dt^2+\frac{r_0^2}{2}H\r^2dx^2+\frac{d\r^2}{\r^2}+d\th^2+\sin^2\th d\ph^2+\cos^2\th d\O_3^2.
\end{align}
where
\begin{align}
f=1-\frac{1}{\r^4},\quad H=1+\frac{1}{\r^4}.
\end{align}
The temperature of the gluon plasma is given by $T=\frac{r_0}{\pi}$.
Note that we set AdS radius $L=1$. It can be reinstated by dimension. We also explicitly factorize $S_5$ into $S_3$ and two
additional angular coordinates $\th$ and $\ph$.
There is also a nontrivial Ramond-Ramond form
\begin{align}
C_4=\(\frac{r_0^2}{2}\r^2H\)^2dt\wg dx_1\wg dx_2\wg dx_3-\cos^4\th d\ph\wg d\O_3.
\end{align}
The D7 branes share the worldvolume coordinates with D3 branes. In
addition, they occupy the coordinates $x_4$-$x_7$ parametrized by the $S_3$ coordinates. Their position in $x_8$-$x_9$ plane can be parametrized by radius $\r\sin\th$ and polar angle $\ph$. The rotational symmetry in the $x_8$-$x_9$ plane corresponds to $U(1)_R$ symmetry in the dual field theory.
The D7 branes has an additional $U(1)_B$ symmetry carried by its worldvolume gauge field. We will use the $U(1)_R$ and $U(1)_B$ symmetries as axial and vector symmetries respectively.

We are interested in the field theory state at finite temperature and finite quark chemical potential $\m_q$ with background magnetic field $\tilde{B}$. To this end, we introduce worldvolume gauge field $A_t(\r)$ and $\tilde{F}_{xy}=\tilde{B}$. 
The embedding function $\th(\r)$ of D7 branes in D3 background is
determined by minimizing the action including a DBI term and WZ term
\begin{align}\label{S_bare}
&S_{D7}=S_{DBI}+S_{WZ}, \no
&S_{DBI}=-N_fT_{D7}\int d^8\x\sqrt{-\text{det}\(g_{ab}+2\pi\a'
  \tilde{F}_{ab}\)}, \no
&S_{WZ}=\frac{1}{2}N_fT_{D7}(2\pi\a')^2\int P[C_4]\wg \tilde{F}\wg \tilde{F}.
\end{align}
Here $T_{D7}$ is the D7 brane tension. $g_{ab}$ and $\tilde{F}_{ab}$ are the
  induced metric and worldvolume field strength respectively. Defining
\begin{align}\label{rescaling}
%&F_{ab}=2\pi\a'\tilde{F}_{ab}, \no
&B=\frac{2\pi\a'}{r_0^2}\tilde{B}, \quad A_t=\frac{2\pi\a'}{r_0}\tilde{A}_t, \no
&\cN=N_fT_{D7}2\pi^2=\frac{N_fN_c\l}{(2\pi)^4},
\end{align}
the action simplifies to
\begin{align}\label{S_redef}
&S_{DBI}=-\frac{\cN}{2\pi^2}\int d^8\x\sqrt{-\text{det}\(g_{ab}+F_{ab}\)}, \no
&S_{WZ}=\frac{1}{4\pi^2}\cN\int P[C_4]\wg F\wg F.
\end{align} 
The asymptotic behavior of $\th$ is given by
\begin{align}\label{mc}
\sin\th=\frac{m}{\r}+\frac{c}{\r^3}+\cdots.
\end{align}
The coefficients are related to bare quark mass $M_q$ and quark condensate $\<\bar{\psi}\psi\>$ \cite{Mateos:2006nu}:
\begin{align}
M_q=\frac{r_0m}{2\pi\a'},\quad \<\bar{\psi}\psi\>=-2\pi\a'\cN r_0^3c.
\end{align}
Similarly, the asymptotic behavior of $A_t$ determines dimensionless quark chemical potential $\mu$ and density $n$:
\begin{align}
A_t=\mu-\frac{n}{\r^2}+\cdots,
\end{align}
with physical quark chemical potential and number density given by $\mu_q=\frac{r_0\mu}{2\pi\a'}$ and $n_q=4\pi\a'\cN r_0^3n$.

The phase diagram of the system has been obtained in \cite{Mateos:2007vc,Kobayashi:2006sb,Filev:2007gb,Erdmenger:2007bn,Evans:2010iy}.
There are two possible embeddings with D7 branes crossing/not crossing
the black hole horizons, corresponding to meson melting/mesonic phase
respectively \cite{Hoyos:2006gb}. We will focus on meson melting phase for studying CSE in quark gluon plasma (QGP). %The possibility of having spatially modulated phase is explored in \cite{}. We will elaborate on this phase and study its correction on CSE.
%Using $t$, $\vec{x}$, $\r$ and angular coordinates on $S_3$ as
%worldvolume coordinates, the induced metric is given by
%\begin{align}\label{ind_metric}
%ds^2_{\text{ind}}=-\frac{r_0^2}{2}\frac{f^2}{H}\r^2dt^2+\frac{r_0^2}{2}H\r^2dx^2+\(\frac{1}{\r^2}+\th'(\r)^2\)d\r^2+\cos^2\th d\O_3^2.
%\end{align}
%We further turn on a constant magnetic field in
%$z$-direction: $F_{xy}=B$, the action of D7 branes can be written as
%\begin{align}
%S_{DBI}=-\cN\int d\r\(\frac{r_0^2}{2}\)^2fH\r^3\sqrt{1+\r^2\th'^2}\sqrt{1+\frac{2B^2}{r_0^2H\r^2}}\cos^3\th,
%\end{align}
%with a vanishing WZ term. The phase diagram in the $m-B$ plane at
%fixed temperature is obtained in Figure.~\ref{}.

\subsection{CSE at finite quark mass}

We consider the fluctuation of embedding function $\ph$ in the above background. The part of quadratic action containing $\ph$ can be written in the following form
\begin{align}\label{compact}
S=\cN\int d^5x\(-\frac{1}{2}\sqrt{-G}G^{MN}\pd_M\ph\pd_N\ph\)
-\cN\k\int d^5x\O\e^{MNPQR}F_{MN}F_{PQ}\pd_R\ph,
\end{align}
with $M=t, x_1, x_2, x_3, \r$. For the evaluation of CSE, we need
\begin{align}
\O=\cos^4\th,\quad \k=\frac{1}{8}.
\end{align}
We do not need explicit form of $G^{MN}$ for now. 
The axial current is defined by \cite{Hoyos:2011us}
\begin{align}\label{JR_def}
J_R^\m=\int d\r \frac{\d {\cal L}}{\d \pd_\m\ph}.
\end{align}
Using EOM of $\ph$, we obtain the following non-conservation equation of axial current
\begin{align}
\pd_\m J^\m_R+\frac{\d {\cal L}}{\d \pd_\r\ph}\vert_{\r=\r_h}^\infty=0.
\end{align}
We will identify $J_R$ as axial current.
The non-conservation of $J_R$
follows from two boundary terms in the integration. The
boundary term at $\r=\infty$ is related to axial anomaly:
\begin{align}\label{Oph}
O_\ph&\equiv -\frac{\d {\cal L}}{\d \pd_\r\ph}\vert_{\r=\infty} \no
&=\cN r_0^4\sqrt{-G}G^{M\r}\pd_M\ph\vert_{\r=\infty}+\k\cN r_0^4\O\e^{MNPQ}F_{MN}F_{PQ}\vert_{\r=\infty} \no
&=O_\h+\cN r_0^4E\cdot B.
\end{align}
Note that the factor $r_0^4$ follows from dimension of ${\cal L}$. In doing this, we have chosen $r_0$ to set unit and work with dimensionless coordinates $t,\,x_1,\,x_2,\,_3$, i.e. $\pd_\m\to r_0\pd_\m$. Combining with \eqref{rescaling}, we obtain $E=\frac{2\pi\a'}{r_0^2}\tilde{E}$ and thus
$\cN EBr_0^4=\frac{N_fN_c}{(2\pi)^2}\tilde{E}\tilde{B}$ corresponding to the anomaly term. Therefore
the term $O_\h$ corresponds to the mass term $iM_q\bar{\psi}\g^5\psi$\footnote{Note that the normalization of $J_R$ is half of $J_5$ in field theory.}. %and $\cN E\cdot B$ corresponds to anomaly term \cite{Guo:2016nnq}.
The other boundary term at horizon $\r=\r_h$ is an artifact of the model. Its presence is tied to our modeling of axial symmetry: since we make use of $U(1)_R$ symmetry for axial symmetry, the gluon plasma is also charged under axial symmetry. The horizon term represents axial charge exchange between the quarks (fundamental matter) and gluons (adjoint matter). The term is indeed non-vanishing in known examples \cite{Hoyos:2011us,Karch:2008uy}. 
However, we will study CSE and correlation functions in static limit. We claim the above artifact is absent in those quantities because charge exchange is not possible in static case. This can be checked explicitly.

\begin{figure}[t]
\includegraphics[width=0.4\textwidth]{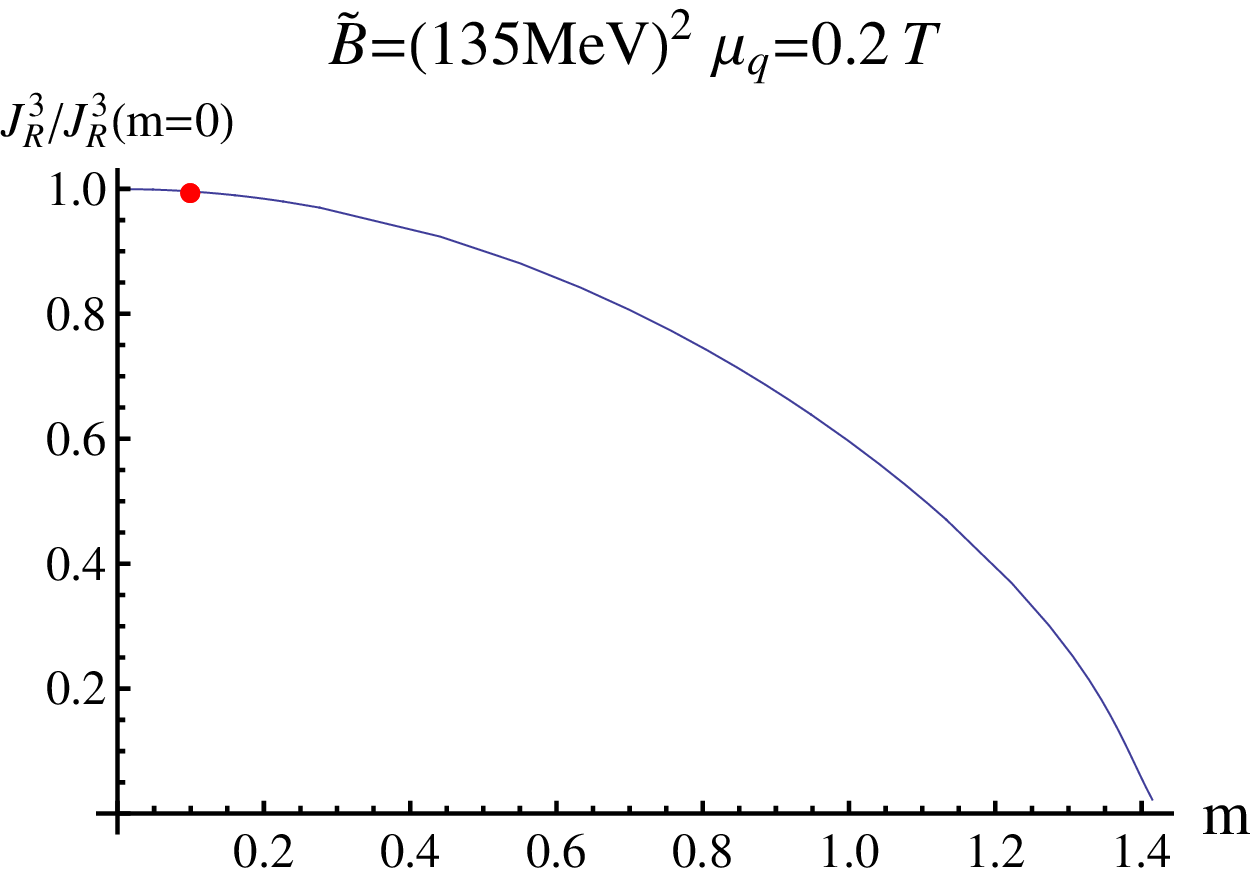}
\includegraphics[width=0.4\textwidth]{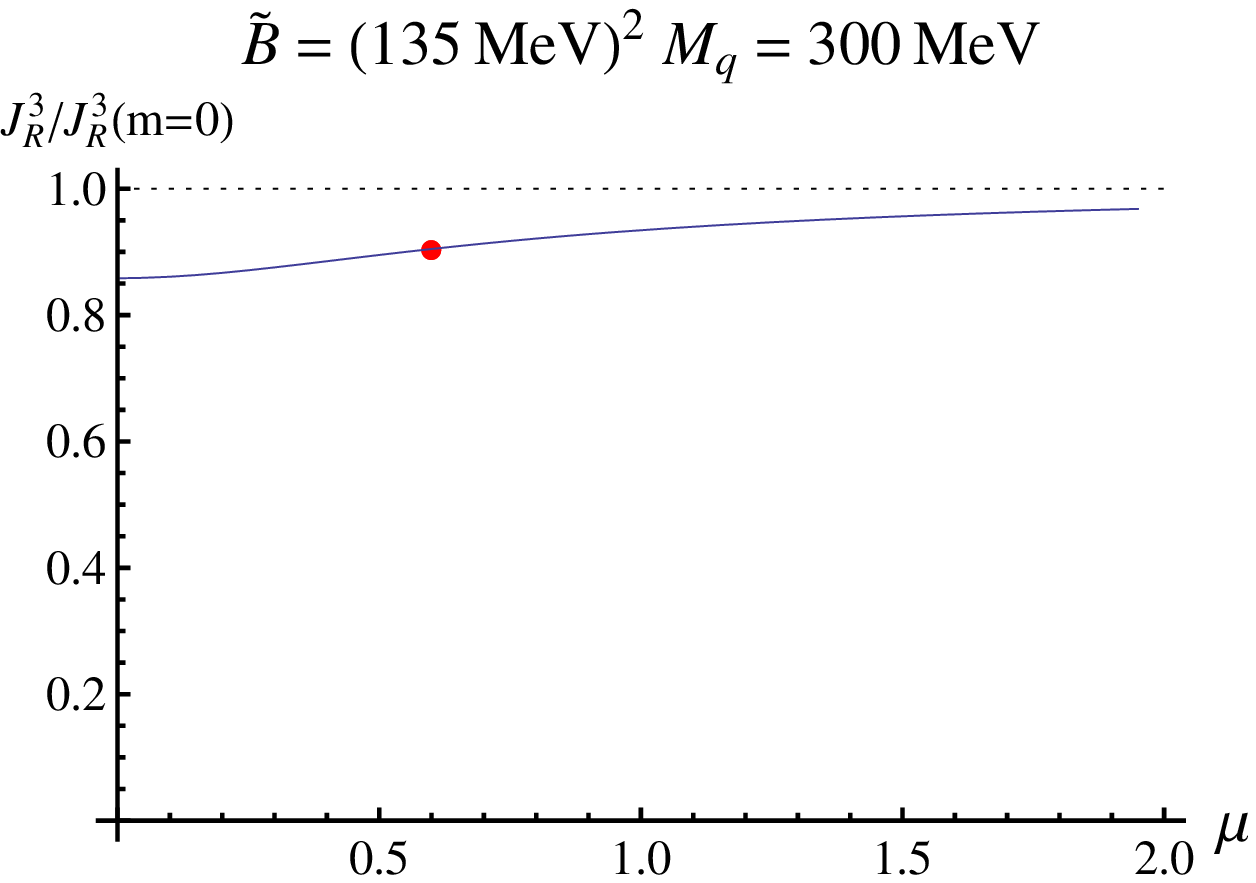}
\includegraphics[width=0.4\textwidth]{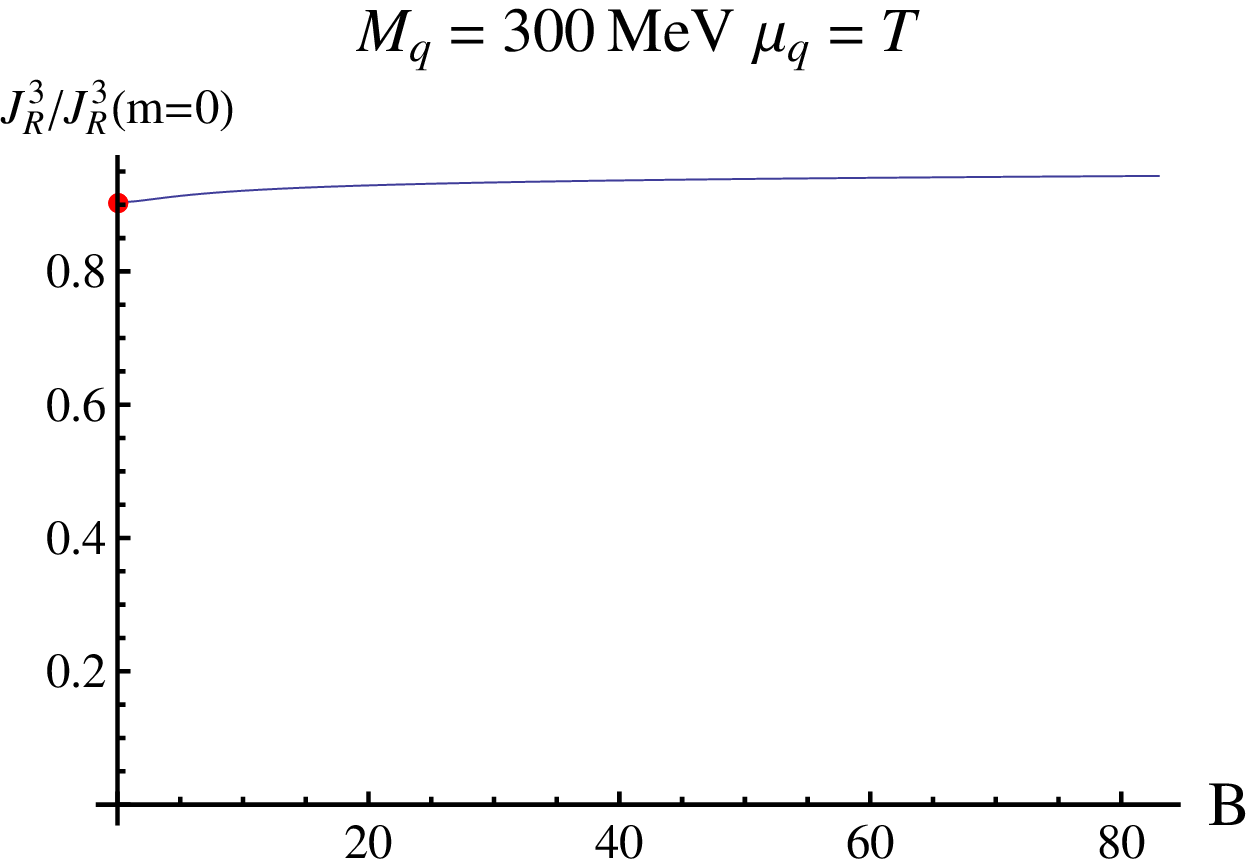}
\caption{\label{fig_j5}normalized $J_R^3$ as a function of $m$, $\m$ and $B$. The temperature of QGP is set to $T=300\MeV$. To guide eyes, we mark phenomenological relevant parameters (strange quark mass $M_q=100\MeV$, $\tilde B=m_{\pi}^2$, $\m_q=0.5T$) with red dots in corresponding panels.
In the upper right and lower panels we use $M_q=300\MeV$ to magnify the dependence on $\m(\mu_q)$ and $B(\tilde{B})$. We use $\a_s=0.3$ in determination of $\l$. For strange quark mass $M_q=100\MeV$, the $\m$ and $B$ dependence is barely visible.}
\end{figure}

Now we proceed to evaluate CSE, which is the axial current $J_R^3$. Note that $\ph=0$ in the background, we obtain
\begin{align}\label{JR_exp}
J_R^3=\cN r_0^3\int_{\r_h}^{\infty} d\r \cos^4\th A_t' B.
\end{align}
We stress that had we assumed $\ph=0$ at the beginning, we would have obtained a vanishing CSE current.
The case $m=0$ is trivial. In this case, the embedding function is given by $\th=0$. $J_R^3$ can be evaluated exactly
\begin{align}
J_R^3=\cN r_0^3\int_{\r_h}^{\infty}A_t' B=\cN r_0^3\mu B=\frac{N_fN_c}{(2\pi)^2}\mu_q\tilde{B}.
\end{align}
This is the standard CSE fixed entirely by anomaly upon restoring units in the last step. Correction to standard CSE exists for $m\ne 0$. In this case, the embedding function $\th$ and gauge potential $A_t$ are only known numerically. The corresponding $J_R^3$ can be obtained by numerical integration. We obtain its dependence on $m$, $\mu$ and $B$ in Figure \ref{fig_j5}. To convert to physical unit, we use
\begin{align}
r_0=\pi T,\quad \a'=\frac{1}{\sqrt{\l}}
%$M_q=\frac{\sqrt{\l}Tm}{2}$, $\mu_q=\frac{\sqrt{\l}T\mu}{2}$ and $\tilde{B}=\frac{\pi\sqrt{\l}T^2B}{2}$
\end{align} 
with phenomenologically relevant coupling and temperature.
We observe that quark mass tends to suppress CSE as expected. Chemical potential and magnetic field both tends to enhance CSE.
The qualitative dependence can be understood from \eqref{j5_corr}. The leading term $-C\m_q\tilde B$ gives the baseline $1$, while the correction is
\begin{align}
\D J_R=\#\frac{M_q^2}{T^2}\m_q\tilde B.
\end{align}
The $m(M_q)$ dependence is quadratic, while the $\m_q$ and $\tilde{B}$ dependence is absent at this order. Assuming higher order terms in $M_q$ can be ignored, the dependence in Fig.\ref{fig_j5} implies the magnitude of the prefactor $\#$ drops with growing $\m_q$ and $\tilde B$. Note that the prefactor is negative, a suppressed magnitude leads to enhancement of CSE.
%We stress that the correction to $J_R^z$ is not due to vev of $im\bar{\psi}\g^5\psi$, which vanishes in the homogeneous background.

\section{Correlators}

In this section, we wish to study the correlator among scalar condensate $\s\equiv{\bar{\psi}}\psi$, pseudoscalar condensate $\s_5\equiv iM_q\bar{\psi}\g^5\psi$ and quark number density $n_q\equiv\bar{\psi}\g^0\psi$. We study the Euclidean correlators at vanishing frequency (in static case when axial charge exchange is absent) and finite momentum.
\begin{align}\label{Gij}
G_{\m\n}(k)=\int d^4(x-y)e^{i\vk\vx}\< O_\m(x)O_\n(y)\>,
\end{align}
with $\m,\n=\s,\,n,\,\s_5$.
To this end, we introduce the following fluctuations to the background:
\begin{align}
&\th(z,\r)=\th(\r)+\d\th(z,\r),\quad A_t(z,\r)=A_t(\r)+a_t(z,\r),\quad \ph=\ph(z,\r).
\end{align}
The open string metric up to quadratic order in fluctuation is given by
\begin{align}
h_{ab}=h_{ab}^{(0)}+h_{ab}^{(1)}+h_{ab}^{(2)}+\cdots,
\end{align}
with
\begin{align}
h_{ab}^{(0)}=
&\begin{pmatrix}
g_{tt}& & -r_0A_t' \\
& g_{xx}& \\
r_0A_t'& & g_{\r\r}+\th'{}^2
\end{pmatrix}\bigoplus
\begin{pmatrix}
g_{xx}& r_0^2B\\
-r_0^2B& g_{xx}
\end{pmatrix}\bigoplus
g_{SS}\bigg(g_{\O3}\bigg), \no
h_{ab}^{(1)}=&\begin{pmatrix}
& -r_0^2\dot{a}_t& -r_0a_t' \\
r_0^2\dot{a}_t& & r_0\d\dot{\th}\th'\\
r_0a_t'& r_0\d\dot{\th}\th'& 2\th'\d\th'
\end{pmatrix}\bigoplus
\bigg(0\bigg)\bigoplus
g_{SS}^{(1)}\bigg(g_{\O3}\bigg), \no
h_{ab}^{(2)}=&\begin{pmatrix}
& &  \\
& r_0^2\(\d\dot{\th}^2+g_{\ph\ph}\dot{\ph}^2\)& r_0\(\d\dot{\th}\d\th'+g_{\ph\ph}\dot{\ph}\ph'\)\\
& r_0\(\d\dot{\th}\d\th'+g_{\ph\ph}\dot{\ph}\ph'\)& \d\th'{}^2+g_{\ph\ph}\ph'{}^2
\end{pmatrix}\bigoplus
\bigg(0\bigg)\bigoplus
g_{SS}^{(2)}\bigg(g_{\O3}\bigg).
\end{align}
Here we use the following coordinates as D7 brane worldvolume coordinates: $t$, $z$, $\r$, $x$, $y$ and $\O$, with $\O$ denoting collectively three angular coordinates on $S_3$. They are ordered as they appear in the open string metric.
It is straight forward but tedious task to work out the quadratic action of the DBI and WZ terms
\begin{align}\label{quad_action}
S_{\text{DBI}}&=-\frac{\cN}{2\pi^2}\int d^8\xi\sqrt{-h}\bigg[\frac{r_0^2}{2}\(\d\dot{\th}^2+g_{\ph\ph}\dot{\ph}^2\)g^{xx}+\frac{1}{2}\(\d\th'^2+g_{\ph\ph}\ph'^2\)h^{\r\r}+\frac{3}{2}\d g_{SS}^{(2)}g^{SS} \no
&+\frac{1}{2}r_0^4\dot{a}_t^2h^{tt}g^{xx}+\frac{1}{2}r_0^2a_t'{}^2h^{tt}h^{\r\r}-\frac{1}{2}r_0^2\(\d\dot{\th}\th'\)^2g^{xx}h^{\r\r}-r_0^3\dot{a}_t\d\dot{\th}\th'h^{t\r}g^{xx} \no
&-r_0a_t'\th'\d\th'h^{t\r}h^{\r\r}-\frac{1}{2}\(\d\th'\th'\)^2h^{\r\r}{}^2 \no
&\frac{3}{8}\d g_{SS}^{(1)}{}^2\(g^{SS}\)^2+\frac{3}{2}\d g_{SS}^{(1)}g^{SS}r_0a_t'h^{t\r}+\frac{3}{2}\d g_{SS}^{(1)}g^{SS}\th'\d\th'h^{\r\r}\bigg], \no
S_{\text{WZ}}&=\frac{\cN}{2\pi^2}r_0^4\int d^8\xi\bigg[\cos^4\th B\(\dot{a}_t\ph'-a_t'\dot{\ph}\)+4\cos^3\th\sin\th\d\th B A_t'\dot{\ph}\bigg].
\end{align}
Here
\begin{align} 
&g_{SS}=\cos^2\th,\;\; \d g_{SS}^{(1)}=-\sin2\th \d\th,\;\; \d g_{SS}^{(2)}=-\cos2\th\d\th^2, \no
&h^{tt}=\frac{g_{\r\r}+\th'{}^2}{g_{tt}\(g_{\r\r}+\th'{}^2\)+r_0^2A_t'{}^2},\;\; h^{\r\r}=\frac{g_{tt}}{g_{tt}\(g_{\r\r}+\th'{}^2\)+r_0^2A_t'{}^2},\;\; h^{t\r}=\frac{-r_0A_t'}{g_{tt}\(g_{\r\r}+\th'{}^2\)+r_0^2A_t'{}^2}, \no
&\int d^8\xi\sqrt{-h}=2\pi^2\int d^5x\sqrt{-\(g_{tt}\(g_{\r\r}+\th'{}^2\)+r_0^2A_t'{}^2\)\(g_{xx}^2+r_0^4B^2\)g_{xx}g_{SS}^3}.
\end{align}
We use dot and prime for derivatives with respect to $z$ and $\r$ respectively. Note that we work with dimensionless z, i.e. $\pd_z\to r_0\pd_z$. This amounts to setting the scale of spatial momentum by temperature. The rescaling makes the $r_0$ dependence of $S_{\text{DBI}}$ and $S_{\text{WZ}}$ appears as an overall $r_0^4$ factor, thus $r_0$ drops out completely from the EOM. %From now on, we will set $r_0=1$ for notationall simplicity and restore $r_0$ when appropriate.
The EOM following from \eqref{quad_action} are given by
\begin{align}\label{eom_fluc}
&\big[2\sqrt{-h}\(3\tan^2\th-\frac{3}{2}\)\d\th-3\sqrt{-h}h^{t\r}\tan\th a_t'+3\(\sqrt{-h}h^{\r\r}\tan\th\th'\)'\d\th+ \no
&\(\sqrt{-h}h^{t\r}h^{\r\r}\th'a_t'\)'-\(\sqrt{-h}h^{\r\r}\(1-\th'{}^2h^{\r\r}\)\d\th'\)'+\sqrt{-h}h^{t\r}\th'g^{xx}\ddot{a}_t \no
&-\sqrt{-h}g^{xx}\(1-h^{\r\r}\th'{}^2\)\ddot{\th}\big]-4\cos^3\th\sin\th BA_t'\dot{\ph}=0, \no
&\big[3\(\sqrt{-h}h^{t\r}\tan\th\d\th\)'-\(\sqrt{-h}h^{tt}h^{\r\r}a_t'\)' +\(\sqrt{-h}h^{t\r}h^{\r\r}\th'\d\th'\)'-\sqrt{-h}h^{tt}g^{xx}\ddot{a}_t+\sqrt{-h}h^{t\r}g^{xx}\th'\d\ddot{\th}\big] \no
&-\(\cos^4\th\)'B\dot{\ph}=0, \no
&\big[\(\sqrt{-h}h^{\r\r}\sin^2\th\ph'\)'+\sqrt{-h}g^{xx}\sin^2\th\ddot{\ph}\big]-\(\cos^4\th\)'B\dot{a}_t-4\cos^3\th\sin\th\d\dot{\th}BA_t'=0.
\end{align}
By observation, we find the ansatz 
\begin{align}\label{ansatz}
\ph(z,\r)=\sin(kz)\ph_k(\r),\quad a_t(z,\r)=\cos(kz)a_t(\r),\quad \d\th(z,\r)=\cos(kz)\d\th(\r), 
\end{align}
solves the $z$-dependence of \eqref{eom_fluc}. To proceed, we note that a generic set of solution has the following asymptotic expansion
\begin{align}
&\ph=f_0+\frac{f_2}{\r^2}+\frac{f_h}{\r^2}\ln\r+\cdots, \no
&a_t=a_0+\frac{a_2}{\r^2}+\frac{a_h}{\r^2}\ln\r+\cdots, \no
&\d\th=\frac{t_1}{\r}+\frac{t_3}{\r^3}+\frac{t_h}{\r^3}\ln\r+\cdots,
\end{align}
where $f_h=-k^2f_0$, $a_h=-k^2a_0$ and $t_h=-k^2t_1$. The leading coefficients are the sources to operators $\s_5$, $\d n$ and $\d\s$ respectively. The subleading coefficients are related to their vevs. A holographic renormalization procedure is needed to determine the vevs. We will elaborate this procedure in appendix A. Here we only show results of correlator $G_{ab}$
\begin{align}\label{dict}
&\frac{G_{\s\s}}{(2\pi\a')^2\cN r_0^2}=\frac{1}{2}S_{\s\s},\quad \frac{G_{nn}}{(2\pi\a')^2\cN r_0^2}=\frac{1}{2}S_{nn},\quad \frac{G_{\s5\s5}}{\cN r_0^4}=\frac{1}{2}S_{\s5\s5}, \no
&\frac{G_{\s n}}{(2\pi\a')^2\cN r_0^2}=\frac{1}{2}S_{\s n},\quad \frac{G_{n \s}}{(2\pi\a')^2\cN r_0^2}=\frac{1}{2}S_{n \s} \no
&\frac{G_{\s \s5}}{(2\pi\a')\cN r_0^3}=\frac{1}{2}S_{\s\s5},\quad \frac{G_{\s5 \s}}{(2\pi\a')\cN r_0^3}=\frac{1}{2}S_{\s5\s}, \no
&\frac{G_{n\s5}}{(2\pi\a')\cN r_0^3}=\frac{1}{2}S_{n\s5},\quad \frac{G_{\s5 n}}{(2\pi\a')\cN r_0^3}=\frac{1}{2}S_{\s5 n},
\end{align}
where we have defined individual responses $S_{ab}$
\begin{align}\label{Ss}
&S_{\s\s}=\frac{\pd t_3}{\pd t_1},\quad S_{\s n}=\frac{\pd t_3}{\pd a_0},\quad S_{\s\s5}=\frac{\pd t_3}{\pd f_0}, \no
&S_{n\s}=-2\frac{\pd a_2}{\pd t_1},\quad S_{n n}=-2\frac{\pd a_2}{\pd a_0},\quad S_{n\s5}=-2\frac{\pd a_2}{\pd f_0}, \no
&S_{\s5\s}=m^2\frac{\pd f_2}{\pd t_1},\quad S_{\s5 n}=m^2\frac{\pd f_2}{\pd a_0},\quad S_{\s5\s5}=m^2\frac{\pd f_2}{\pd f_0}.
\end{align}
%%%%%%%%%%%%%%%%%%%%%%%%%%%%%%%%
We proceed to solve \eqref{eom_fluc}. Since we have three coupled differential equations, we expect to have three independent solutions. We solve \eqref{eom_fluc} by numerical integration from the horizon to the boundary. The initial condition we impose at the horizon is regularity condition. In practice, we start off the horizon with the following three independent solutions:
\begin{align}
&\d\th^{(1)}(1+\e)=1+O(\e^2),\;\; a_t^{(1)}=O(\e^2),\;\; \ph^{(1)}=O(\e^2), \no
&\d\th^{(2)}(1+\e)=O(\e^2),\;\; a_t^{(2)}=\e^2+O(\e^3),\;\; \ph^{(2)}=O(\e^2), \no
&\d\th^{(3)}(1+\e)=O(\e^2),\;\; a_t^{(3)}=O(\e^2),\;\; \ph^{(3)}=1+O(\e^2).
\end{align}
These solutions give rise to the following asymptotics at the boundary
\begin{align}\label{solutions}
&\ph^{(i)}=f_0^{(i)}+\frac{f_2^{(i)}}{\r^2}+\frac{f_h^{(i)}}{\r^2}\ln\r+\cdots, \no
&a_t^{(i)}=a_0^{(i)}+\frac{a_2^{(i)}}{\r^2}+\frac{a_h^{(i)}}{\r^2}\ln\r+\cdots, \no
&\d\th^{(i)}=\frac{t_1^{(i)}}{\r}+\frac{t_3^{(i)}}{\r^3}+\frac{t_h^{(i)}}{\r^3}\ln\r+\cdots,
\end{align}
with $i=1,2,3$ labeling different solutions.
In order to calculate individual responses $S_{ab}$, we need to construct proper solution for which the other two sources vanish. This can be done efficiently in the following way
\begin{align}\label{matrix3}
\begin{pmatrix}
S_{\s\s}& S_{n\s}& S_{\s5\s}\\
S_{\s n}& S_{nn}& S_{\s5n}\\
S_{\s\s5}& S_{n\s5}& S_{\s5\s5}
\end{pmatrix}=
\begin{pmatrix}
t_1^{(1)}& a_0^{(1)}& f_0^{(1)}\\
t_1^{(2)}& a_0^{(2)}& f_0^{(2)}\\
t_1^{(3)}& a_0^{(3)}& f_0^{(3)}
\end{pmatrix}^{-1}
\begin{pmatrix}
t_3^{(1)}& -2a_2^{(1)}& m^2f_2^{(1)}\\
t_3^{(2)}& -2a_2^{(2)}& m^2f_2^{(2)}\\
t_3^{(3)}& -2a_2^{(3)}& m^2f_2^{(3)}
\end{pmatrix}.
\end{align}
On general ground, we expect the Euclidean correlator to be real and symmetric $G_{ab}=G_{ab}^*=G_{ba}$. Our numerical results confirm that this is indeed the case. 
We also find the following scaling of all individual responses at small $k$.
\begin{align}\label{scaling}
&S_{\s\s},\,S_{\s n},\,S_{n\s},\,S_{nn}\sim O(k^0), \no
&S_{\s\s5},\,S_{n\s5},S_{\s5\s},\,S_{\s5n},\sim O(kB), \no
&S_{\s5\s5}\sim O(k^2)
\end{align}
The first line of \eqref{scaling} is closely related to thermodynamics of the system. %At $k=0$, we find $S_{\s n}$ and $S_{n\s5}$ are equal to each other within numerical accuracy. In fact, this is not a coincidence. 
At $k=0$, we have
\begin{align}
S_{\s n}=\frac{\pd c(m,\,\m)}{\pd \m},\quad S_{n\s}=\frac{\pd n(n,\,\m)}{\pd m}.
\end{align}
Similarly, the diagonal responses $S_{\s\s}$ and $S_{nn}$ are related to $\frac{\pd c}{\pd m}$ and $\frac{\pd n}{\pd \m}$\footnote{For $S_{\s\s}$, there is correction proportional to $m^2$}. We have compared the results of individual responses at $k=0$ and those of thermodynamics, finding expected agreement.

The second line of \eqref{scaling} is of more interest to us. It follows from \eqref{dict} that $G_{\s\s5},\,G_{n\s5}\sim O(kB)$. This is consistent with parity (P) and time-reversal (T) symmetry of the corresponding operators: $\s_5$ is odd under both P and T, while $\s$ and $n$ are even under $P$ and $T$. The external $B$ and momentum $k$ are odd under $T$ and $P$ respectively.
These Euclidean correlators characterize response of the system to external parameter $\ph$:
\begin{align}
\s\sim G_{\s\s5}\ph,\quad n\sim G_{n\s5}\ph.
\end{align}
Here we use $\ph$ to denote the field theory source coupled to $\s_5$. The scaling $O(k)$ can be understood as follows: the source $\ph$ enters field theory Lagrangian as $M_q\bar{\ps}e^{i\ph\g^5}\ps$. If we perform a chiral rotation, $\ps\to e^{-i\g^5\ph/2}\ps$, the relevant terms in the Lagrangian is modified as \cite{Hoyos:2011us}
\begin{align}
M_q\bar{\ps}e^{i\ph\g^5}\ps\to M_q\bar{\ps}\ps-\frac{\pd_\m\ph}{2}\bar{\ps}\g^\m\g^5\ps.
\end{align}
This implies that only $\pd_\mu\ph$ appears as physical parameter. In our case, $\ph$ only depends on $z$ in field theory coordinates, therefore the physical parameter is $\dot{\ph}$. Interestingly, $\dot{\ph}$ can be identified as the chiral shift parameter proposed in \cite{Gorbar:2011ya,Gorbar:2013upa}. Assuming the response of $\s$ and $n$ to $\dot{\ph}$ is $O(1)$ at small $k$, we naturally explain the $O(k)$ scaling of correlators $G_{\s\s5},\,G_{n\s5}$.

The third line of \eqref{scaling} indicates $G_{\s5\s5}\sim O(k^2)$. $G_{\s5\s5}$ is by definition the susceptibility of $\s_5$. The susceptibility $G_{\s5\s5}$ is parity even, thus it scales as even power of $k$. As we argued above, any response to $\ph$ has to start from $O(k)$, the most probable scaling is $O(k^2)$.

\begin{figure}[t]
\includegraphics[width=0.5\textwidth]{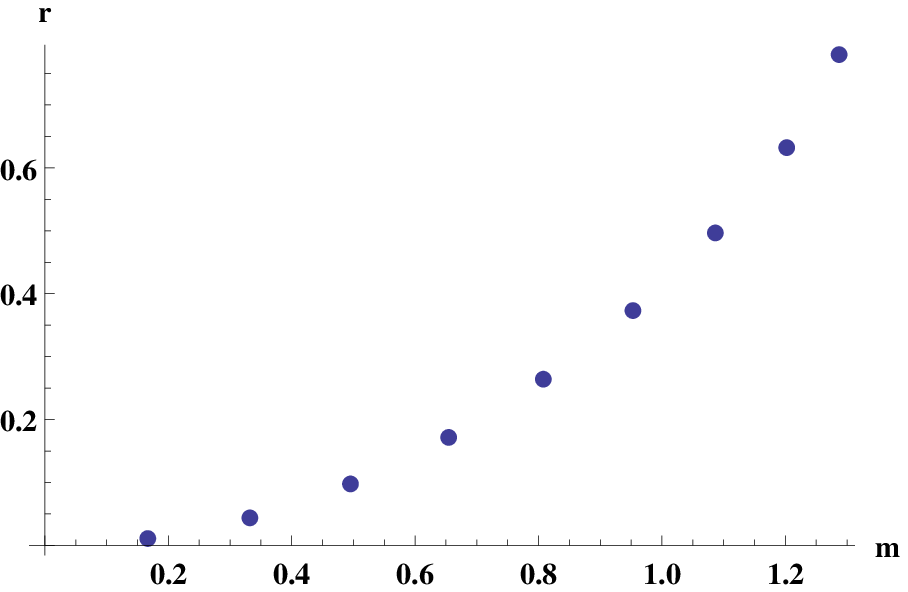}
\includegraphics[width=0.5\textwidth]{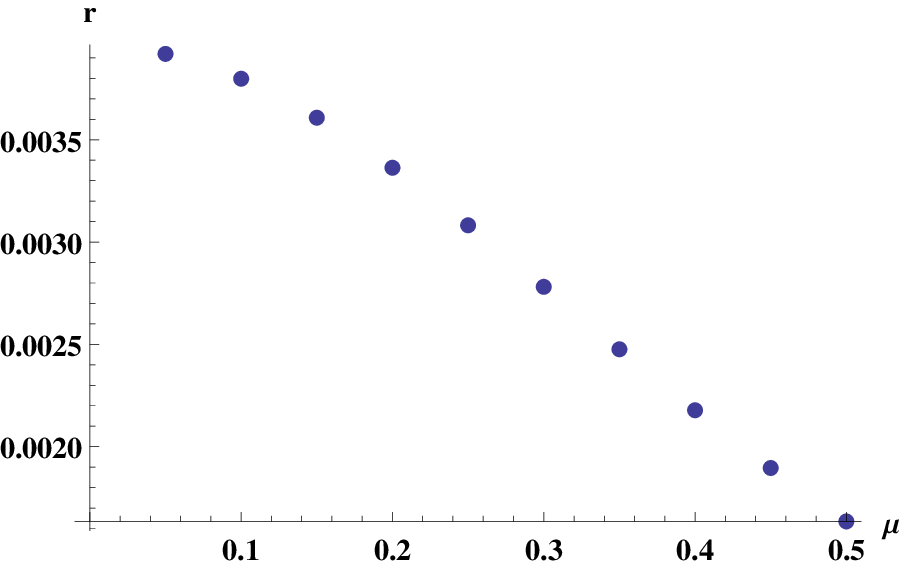}
\caption{\label{fig_r}$r$ defined in \eqref{r_def} as a function of $m$ at $\m=0$ (left). Note that $m$ is related to physical quark mass $M_q$ by $m=\frac{2}{\sqrt{\l}}\frac{M_q}{T}$. The right plot shows $r$ as a function of $\m$ at $m=0.1$. The ratio $r$ increases with $m$, but decreases with $\m$. %We have used $\a_s=0.3$ in determination of $\l$. Our holographic result is shown in points. The result from a Wigner function calculation is shown in curve as a comparison [Qun Wang, private communication].
}
\end{figure}

Let us take a closer look at correlators involving $\s_5$. We will present results primarily on these correlators at small $k$. In fact, we can confirm the linear scaling relation described above by perturbative calculation in $k$. Note that $\ph\sim O(k)$, $a_t,\,\d\th\sim O(1)$. To the lowest non-trivial order in $k$, we only need to solve the following equations
\begin{align}\label{k_eom}
&2\sqrt{-h}\(3\tan^2\th-\frac{3}{2}\)\d\th-3\sqrt{-h}h^{t\r}\tan\th a_t'+3\(\sqrt{-h}h^{\r\r}\tan\th\th'\)'\d\th+ \no
&\(\sqrt{-h}h^{t\r}h^{\r\r}\th'a_t'\)'-\(\sqrt{-h}h^{\r\r}\(1-\th'{}^2h^{\r\r}\)\)'=0, \no
&3\(\sqrt{-h}h^{t\r}\tan\th\d\th\)'-\(\sqrt{-h}h^{tt}h^{\r\r}a_t'\)' +\(\sqrt{-h}h^{t\r}h^{\r\r}\th'\d\th'\)'=0, \no
&\(\sqrt{-h}h^{\r\r}\sin^2\th\ph'\)'-\(\cos^4\th\)'B\dot{a}_t-4\cos^3\th\sin\th\d\dot{\th}BA_t'=0.
\end{align}
From the first two equations of \eqref{k_eom}, we can solve for $\d\th$ and $a_t$. Plugging the solution into the third equation and integrating from the horizon to the boundary, we obtain
\begin{align}\label{sigma5}
\sqrt{-h}h^{\r\r}\sin^2\th\ph'\vert_{\r_h}^\infty=\int_{\r_h}^\infty d\r\(\cos^4\th\)'B\dot{a}_t+4\cos^3\th\sin\th\d\dot{\th}BA_t'
\end{align}
On the left hand side (LHS), the boundary term at the horizon vanishes, the boundary term at infinity is just $-\frac{m^2f_2}{2}$. On the right hand side (RHS), it is related to the sources $t_1$ and $a_0$.
There are two independent solutions. We denote their asymptotics as
\begin{align}
&\d\th^{(i)}=\frac{t_1^{(i)}}{\r}+\frac{t_3^{(i)}}{\r^3}+\frac{t_h^{(i)}}{\r^3}\ln\r+\cdots, \no
&a_t^{(i)}=a_0^{(i)}+\frac{a_2^{(i)}}{\r^2}+\frac{a_h^{(i)}}{\r^2}\ln\r+\cdots,
\end{align}
with $i=1,2$. Using \eqref{sigma5}, each solution give rise to vev of $\s_5\sim m^2f_2^{(i)}$. Similar to \eqref{matrix3}, we obtain the correlators as
\begin{align}\label{matrix2}
\begin{pmatrix}
S_{\s5\s}\\
S_{\s5n}
\end{pmatrix}=
\begin{pmatrix}
t_3^{(1)}& a_0^{(1)}\\
t_3^{(2)}& a_0^{(2)}
\end{pmatrix}^{-1}
\begin{pmatrix}
m^2f_2^{(1)}\\
m^2f_2^{(2)}
\end{pmatrix}.
\end{align}
Note that at the boundary $\dot{a}_t\sim E$, $\d\dot{\th}\sim \d\dot{m}$. The correlator $S_{\s5n}(G_{\s5n})$ measures the response of $\s_5$ to parallel $E$ and $B$ fields. To study the response in more detail, we define a dimensionless ratio
\begin{align}\label{r_def}
r=-\frac{\s_5}{\cN E\cdot B r_0^4}
\end{align}
Note that we have included a minus sign in the definition of $r$ such that $r$ is always positive. 
In terms of correlators, $r=\frac{S_{\s5n}}{2ik}$. Since $S_{\s5n}\sim O(k)$, $r$ approaches a constant in the limit $k\to0$.
We plot the $m$ and $\m$-dependence of $r(k\to0,\m=0)$ in Figure \ref{fig_r}. We find $r$ increases with $m$, but decreases with $\m$. The dependence is in qualitative agreement with our discussion before: $r\sim g\sim \#\frac{M_q^2}{T^2}$, which grows with $m$, and drops with $\m_q$ from the $\m_q$ dependence of the prefactor $\#$.
Now we are in a position to confirm the claim \eqref{j5_corr} in Sec I.
We begin by working in the background with $\m=0\,(A_t=0)$. The corresponding correlator $G_{\s5n}$ becomes particularly simple then
\begin{align}\label{Gs5nk0}
G_{\s5n}(k\to0)=-\(2\pi\a'\)\cN r_0^3\frac{\int_{\r_h}^\infty d\r\(\cos^4\th\)'B\dot{a}_t}{a_t(\r\to\infty)}.
\end{align}
Noting that $\frac{r_0\m}{2\pi\a'}=\m_q$, we obtain
\begin{align}\label{Gmq}
\s_5=G_{\s5n}(k\to0,\m=0)\m_q=-i\cN r_0^4 k\int_{\r_h}^\infty\(\cos^4\th\)'a_tB.
\end{align}
One the other hand, the correction to CSE can be obtained by performing an integration by part on \eqref{JR_exp}
\begin{align}\label{jr_split}
J_R^3&=\cN r_0^3\int d\r \cos^4\th A_t'B=\cN r_0^3\(\cos^4\th A_t'B\vert_{\r_h}^\infty-\int_{\r_h}^\infty\(\cos^4\th\)'A_tB\) \no
&=\cN r_0^3\(\m B-\int_{\r_h}^\infty\(\cos^4\th\)'A_tB\).
\end{align}
The first term of \eqref{jr_split} corresponds to the standard CSE, while the second term comes from mass correction, which is precisely \eqref{Gmq}.
%Comparing with \eqref{Gs5nk0}, we easily establish a relation between the second term and $G_{\s5n}$
%Now we look at the correlator $S_{\s5n}$ in more detail.
%Note that the LHS of \eqref{Gmq} is just $\s_5$.
%A physical measure of $\s_5$ is the ratio of $\s_5$ vev and external $E\cdot B$ field:
In this case, $r(k\to0)$ takes the form
\begin{align}
r(k\to0)=\frac{\int_{\r_h}^\infty d\r\(\cos^4\th\)'{a}_t}{a_t(\r\to\infty)}.
\end{align}
It is instructive to analyze the coupling dependence of $r$:
we first note that in the case $\m=0$, the coupling enters only through $m=\frac{2}{\sqrt{\l}}\frac{M_q}{T}$. On the other hand, we have argued in the introduction that $r\sim g\sim \#\frac{M_q^2}{T^2}$, which suggests the following dependence $r\sim \frac{1}{\l}\frac{M_q^2}{T^2}$. This seems to imply that stronger interaction leads to weaker response of $\s_5$ to external electromagnetic fields. This interpretation is misleading for the following reason: in D3/D7 model, the electromagnetic coupling to quark is the same as strong coupling, thus $\cN EB\sim O(\l)$, so the actual response of $\s_5$ is $O(\l^0)$. %A weak coupling approach based on Wigner function shows that $r\sim O(\l^0)$ \cite{}.
It is also interesting to compare $r$ with the same quantity studied in \cite{Guo:2016nnq}, which is defined in the regime $\o\to0,\, k=0$. In fact, we can show analytically that they do not agree. For monotonic $a_t$, we have
\begin{align}
r(k\to0,\,\o=0)=\frac{\int_{\r_h}^\infty d\r\(\cos^4\th\)'B{a}_t}{a_t(\r\to\infty)}<\int_{\r_h}^\infty d\r\(\cos^4\th\)'B&=\(1-\cos^4\th_h\)B \no
&=r(\o\to0,\, k=0).
\end{align}
This reveals noncommutativity of the limits $\o\to0,\,k\to0$ and $k\to0,\,\o\to0$ in the response of $\s_5$.
The correlator $G_{\s5n}$ tells us more than the response of $\s_5$. Note that we have $G_{\s5n}=G_{n\s5}$ by symmetry. $G_{n\s5}$ characterizes the response of $n$ to chiral shift $\dot\ph$. The result of $r$ indicates that chiral shift can also induce correction to $n$ in the presence of $B$, with the correction increases with $m$, but decreases with $\m$.

\begin{figure}[t]
\includegraphics[width=0.5\textwidth]{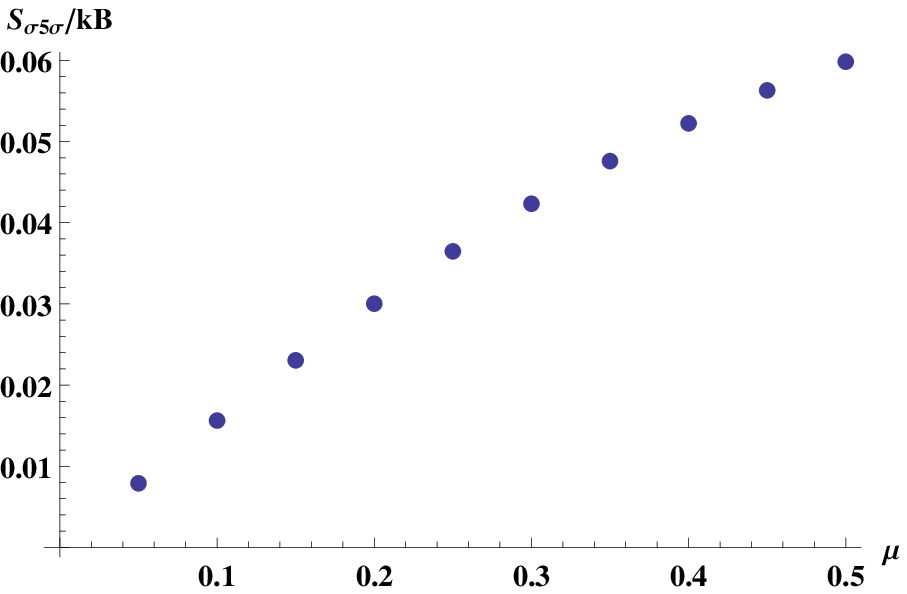}
\includegraphics[width=0.5\textwidth]{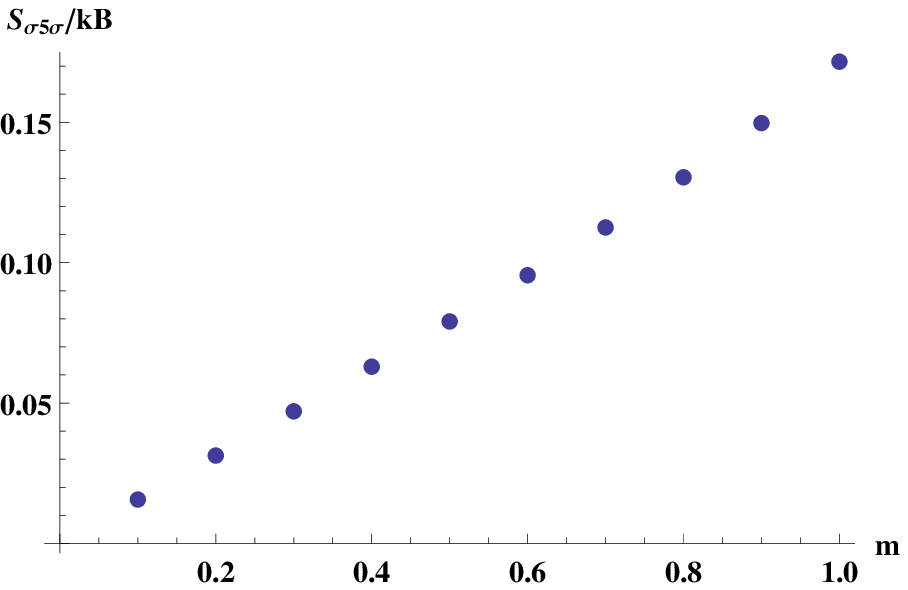}
\caption{\label{fig_Gs5s}$\lim_{k\to0}\frac{S_{\s5\s}}{Bk}$ as a function of $\m$ at $m=0.1$ (left) and the same quantity as a function of $m$ at $\m=0.1$ (right). The response of $\s_5$ to spatially varying mass increases with both $\m$ and $m$.}
\end{figure}

Now we turn to $S_{\s5\s}$. This correlator measures the response of $\s_5$ to spatially varying quark mass $\d m$. %It can be understood in the following way. The standard CSE result $j_5=\cN \sqrt{\m^2-m^2}B$. It suggests an effective chemical potential $\sqrt{\m^2-m^2}$ for massive quark. It follows that spatially varying quark mass induces spatially varying effective chemical potential, which generates effective electric field. 
We plot the $m$-dependence and $\m$-dependence of $S_{\s5\s}$ in Figure \ref{fig_Gs5s}.
Indeed, we can see in Fig.~\ref{fig_Gs5s} that $G_{\s5\s}$ vanishes approximately linearly in $\m$ and $m$, which is clear evidence for \eqref{s5_dm}.
By symmetry $S_{\s5\s}=S_{\s\s5}$, we also obtain that chiral shift $\dot{\ph}$ can induce correction to $\s$. The correction increases with both $\m$ and $m$.
%The claim we made in the introduction that $\s_5$ links both $S_{\s5n}$ and $S_{\s5\s}$ is clearly evidenced in Figure \ref{fig_SS}.
%\begin{figure}[t]
%\includegraphics[width=0.5\textwidth]{SSratio}
%\caption{\label{fig_SS}$\frac{mS_{\s5s}}{2\m G_{\s5n}}$ as a function of $m$ at $\m=0.1$. The ratio is numerically consistent with unity for small $m$ and shows deviation as $m$ increases.}
%\end{figure}

\begin{figure}[t]
\includegraphics[width=0.5\textwidth]{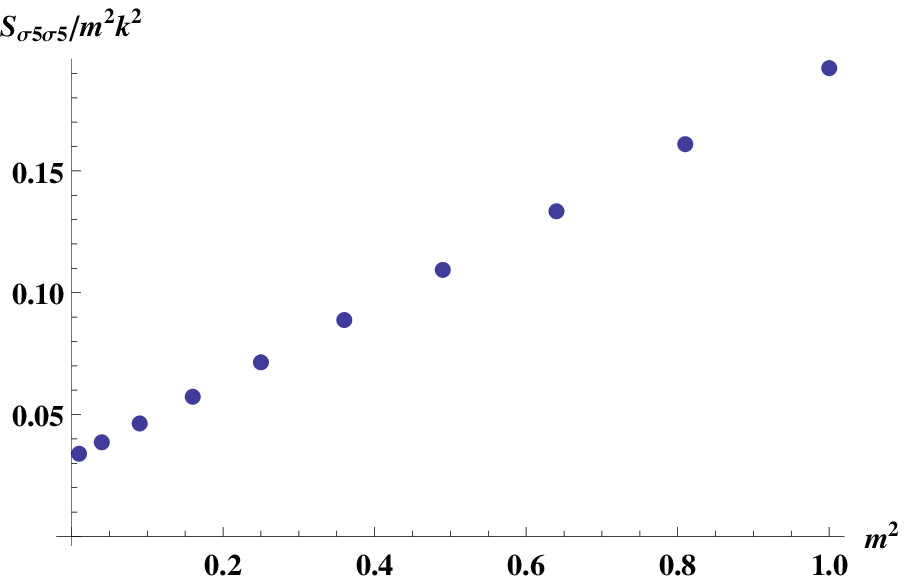}
\includegraphics[width=0.5\textwidth]{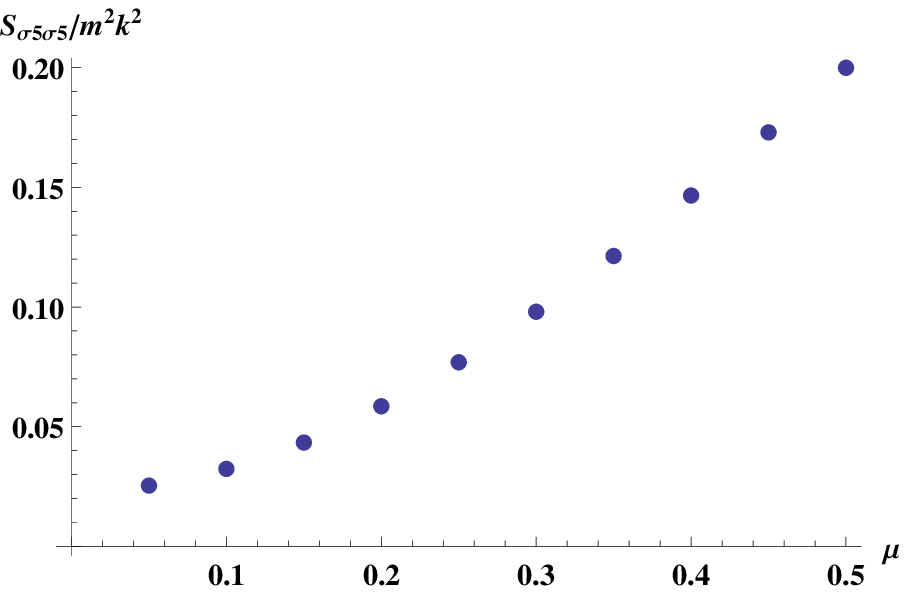}
\caption{\label{fig_Gs5s5}$\lim_{k\to0}\frac{S_{\s5\s5}}{m^2k^2}$ as a function of $m$ at $\m=0.1$ (left) and the same quantity as a function of $\m$ at $m=0.1$ (right). The left plot is suggestive of the expansion $\lim_{k\to0}\frac{S_{\s5\s5}}{m^2k^2}=a+bm^2+o(m^2)$. The right plot shows the response of $\s_5$ to $\ph$ increases with $\m$.}
\end{figure}

Finally we plot the $m$ and $\m$ dependence of $S_{\s5\s5}$ in Figure \ref{fig_Gs5s5}. The scaling $S_{\s5\s5}\sim O(k^2)$ allows for the following parametrization of $\s_5$
\begin{align}\label{para}
\nabla\cdot{\bf j}_5=2\s_5=h(m^2,\m,B)\nabla^2\ph.
\end{align}
From \eqref{para}, we easily obtain an induced $j_5$ in the presence of chiral shift $\dot{\ph}$:
\begin{align}
{\bf j}_5=h(m^2,\m,B)\nabla\ph.
\end{align}
A similar current from spatial gradient of axion is also discussed in \cite{Iatrakis:2014dka}.
As we discussed before, $\nabla\ph$ is just the chiral shift parameter, which couples to the axial current in the Lagrangian. The function $h$ can be viewed as an effective susceptibility.
Fig.\ref{fig_Gs5s5} suggests the following dependence $\frac{h}{\cN}=am^2+bm^4$. Converting to physical parameters, we have $h=\#M_q^2+O(M_q^2)$, with $\#\sim O(\l^0)$.

\section{Normalizable mode}

Now we extend our study to correlators in regime of arbitrary $k$. Instead of calculating all components of correlators, we look for normalizable modes. The existence of normalizable mode means that it costs no energy to support such a mode. It usually corresponds to spontaneous generation of spiral phase with the spatial period set by the momentum of the normalizable mode. The normalizable mode corresponds to the point where the determinant vanishes:
\begin{align}
\begin{vmatrix}
t_1^{(1)}& a_0^{(1)}& f_0^{(1)}\\
t_1^{(2)}& a_0^{(2)}& f_0^{(2)}\\
t_1^{(3)}& a_0^{(3)}& f_0^{(3)}
\end{vmatrix}
=0
\end{align}

\begin{figure}[t]
\includegraphics[width=0.5\textwidth]{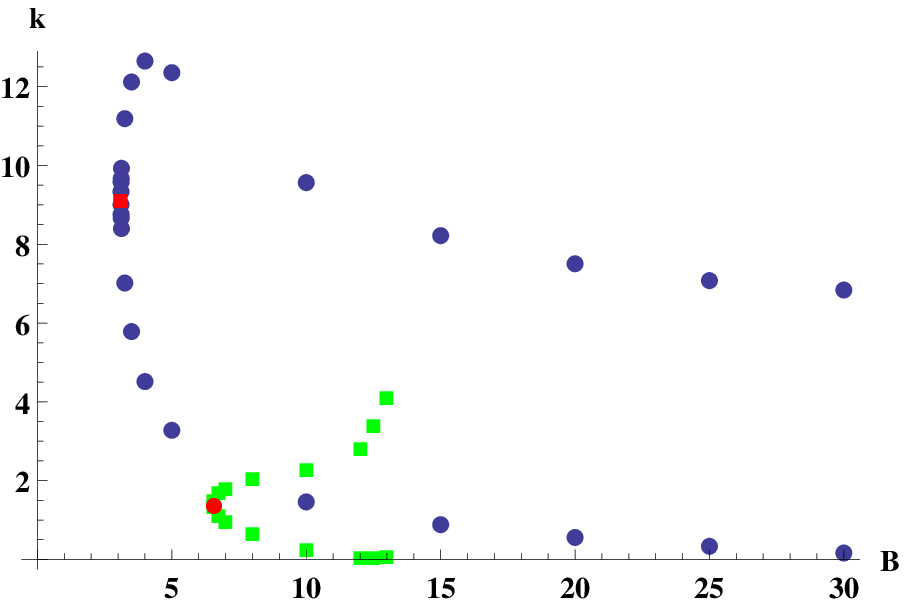}
\includegraphics[width=0.5\textwidth]{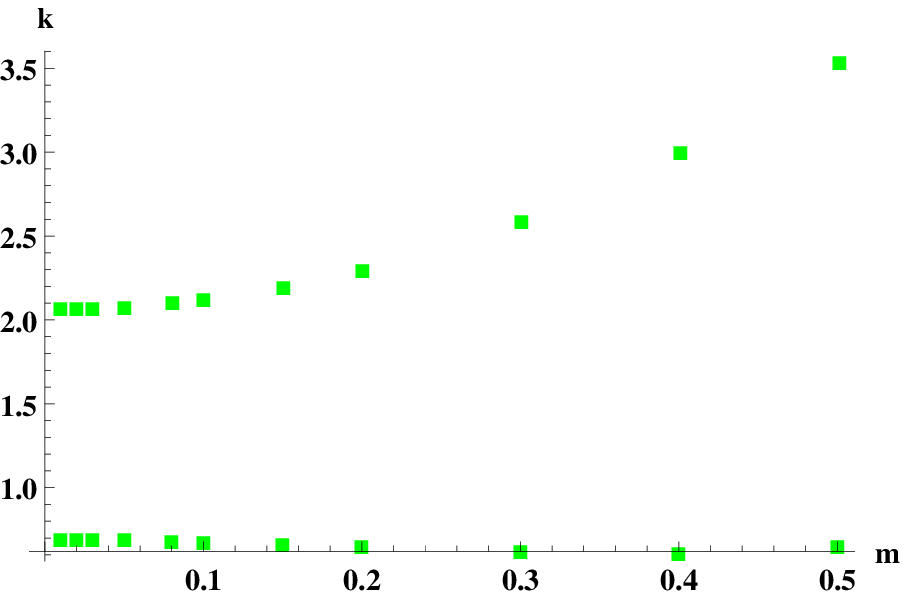}
\caption{\label{fig_kB}Momentum of normalizable mode as a function of $B$ at $m=0.05$ (left). The normalizable modes appear in a pair for each $B$, giving rise to two banches. The blue dots and green squares correspond to the cases with $\mu=3$ and $\mu=1$ respectively. There is a critical $B_c$ marked by red dots (squares) for each case, corresponding to the point where two momenta merge. The normalizable modes start to appear with finite $k$ beyond $B_c$, indicating a first order transition. At larger $B$, the state possibly become metastable. Momentum of normalizable mode as a function of $m$ for $\mu=1$ and $B=8$ (right).}
\end{figure}

We show the momentum $k$ of the mode as a function of $B$ in Figure \ref{fig_kB}.
We find that normalizable modes exist for medium with general nonvanishing $\m$ beyond certain critical magnetic field $B_c$. For each $B>B_c$, there are two normalizable modes with different momenta $k$. The low momentum branch appears monotonic decreasing function of $B$, while the high momentum branch is non-monotonic. The normalizable modes we find are numerically consistent with the quasi-normal mode reported in \cite{Kharzeev:2011rw}. 
It is interesting to note that the critical magnetic field corresponds to the point where the two momenta merge. Furthermore, the modes extend to the region of large $B$, where the state possibly becomes metastable \cite{Evans:2010iy}. We do not keep the corresponding mode in Fig.~\ref{fig_kB}. 
Turning to the $\m$ dependence, we see that as $\m$ is lowered, $B_c$ grows. This is qualitatively in agreement with the chiral soliton solution found in \cite{Brauner:2016pko} in confined phase.
We also show $k$ as a function of $m$ in Fig.~\ref{fig_kB}, which clearly shows the low/high momentum branches. As a function of $m$, the high momentum branch appears monotonic increasing function, while the low momentum branch is non-monotonic.

To have an idea on the magnitude of magnetic field, we convert ${B}_c$ for the case $\m=3$ to physical unit. For gluon plasma at temperature $T=300\MeV$ and coupling $\a_s=0.3$. This correspond to $\tilde{B}_c=(389\MeV)^2$ for $\m_q=504\MeV$. 
%We find for sufficient large $B$ (still within the meson mellting phase), a normalizable mode always exists. The normalizable mode is numerically consistent with one of the unstable quasi-normal mode found in a previous study \cite{Kharzeev:2011rw}. For small $B$, the instability disappears, indicating the corresponding transition is first order. We plot the dependence of momentum of the mode on the magnetic field in Figure \ref{fig_kB}.

\section{Outlook}

We close this paper by discussing several open questions that we may address based on the results of this paper. Firstly, how does quark mass affect the dynamics of axial and vector charge. As we have seen the pseudoscalar condensate responses to gradient of quark chemical potential etc. This brings in an additional coupling between axial and vector charges. Consequently, it should also modify the dispersion of CMW. For strange quark mass, we expect from the Fig.~\ref{fig_j5} that the modification in phenomenology is modest. Since our study focuses on Euclidean correlators, the same quantities can be reliably studied on the lattice, which will provide quantitative answers for quark mass effect in real world QCD.

Secondly, the normalizable mode we found at sufficient large $B$ and $\m$ suggests possible formation of spiral phase. To find the true ground state, we need to go beyond the linear analysis. We expect that the true ground state is characterized by the spontaneous generation of chiral shift, which induces further correction to axial current. The correlator $G_{\s\s5}$ and $G_{n\s5}$ indicates the correction to $\s$ and $n$ as well. It would be interesting to find out detail about this state. We leave this for future work.

Last but not the least, the normalizable mode provides an explicit example of spiral phase in meson melting phase. It would be interesting to extend this work to mesonic phase. Such an example has been found at zero temperature in \cite{Brauner:2016pko} based on effective field theory models. It would be interesting to study the stability of such state against finite temperature fluctuation.

\section*{Note added}
When this work was near complete, we learned that Qun Wang et al was about to finish a closely related work \cite{Fang:2016uds}. We thank Qun Wang for sharing with us notes of their work before publication.

\section*{Acknowledgments}

We are grateful to Gao-Qing Cao, Kenji Fukushima, Xu-Guang Huang, Jinfeng Liao, Yin Jiang, Qun Wang and Yi Yin for useful discussions and correspondence. S.L. is partially supported by Junior Faculty's Fund of Sun Yat-Sen University and One Thousand Talent Program for Young Scholars. He also thanks Central China Normal University for hospitality during the workshop ``QCD Phase Structure III'', where part of this work was done.

\appendix

\section{Dictionary for Euclidean Correlator}

To find out correlators $G_{ab}$, we first obtain on-shell action from \eqref{eom_fluc}:
\begin{align}\label{S_pd}
S^\pd_{\r=\L}=&-\cN r_0^4\int d^4x\sqrt{-h}\big[\frac{1}{2}\(\d\th\d\th'\(1-\th'{}^2h^{\r\r}\)+g_{\ph\ph}\ph\ph'\)h^{\r\r}+\frac{1}{2}a_ta_t'h^{tt}h^{\r\r}+\frac{3}{4}\d g_{SS}^{(1)}g^{SS}\th'\d\th h^{\r\r} \no
&-\frac{1}{2}a_t\th'\d\th'h^{t\r}h^{\r\r}-\frac{1}{2}a_t'\th'\d\th h^{t\r}h^{\r\r}+\frac{3}{4}\d g_{SS}^{(1)}g^{SS}a_th^{t\r}\big] \no
&+\frac{1}{2}\cos^4\th B\(\dot{a}_t\ph-a_t\dot{\ph}\).
\end{align}
The asymptotics of fields can be obtained from EOM as
\begin{align}\label{asym_sol}
&\ph=f_0+\frac{f_2}{\r^2}+\frac{f_h}{\r^2}\ln\r+\cdots, \no
&a_t=a_0+\frac{a_2}{\r^2}+\frac{a_h}{\r^2}\ln\r+\cdots, \no
&\d\th=\frac{t_1}{\r}+\frac{t_3}{\r^3}+\frac{t_h}{\r^3}\ln\r+\cdots,
\end{align}
with the coefficient of logarithmic terms fixed as
\begin{align}\label{log_coeff}
f_h=-k^2f_0,\quad a_h=-k^2a_0,\quad t_h=-k^2t_1.
\end{align}
Plugging \eqref{asym_sol} into \eqref{S_pd}, we find the second line always vanishes in the limit $\L\to\infty$. The first line gives the following contribution
\begin{align}\label{reg}
S^{\pd}&=-\cN r_0^4\big[-\frac{t_1^2\L^2}{8}+\frac{1}{8}\(6m^2t_1^2-4t_1t_3+t_1t_h-4t_1t_h\ln\L\) \no
&+\frac{1}{4}\(2a_0a_2-a_0a_h+2a_0a_h\ln\L\)+\frac{m^2}{8}\(-2f_0f_2+f_0f_h-2f_0f_h\ln\L\)\big]+\cdots
\end{align}
We need the following counter terms to remove quadratic divergence in \eqref{reg}
\begin{align}\label{counter}
S^{\text{counter}}_{\r=\L}=\frac{1}{2}\cN\sqrt{-\g}\d\th^2=\cN\big[\frac{t_1^2\L^2}{8}+\frac{1}{4}\(t_1t_3+t_1t_h\ln\r\)\big]+\cdots.
\end{align}
The coefficient of the logarithmic terms \eqref{log_coeff} is a special case of a more general relation:
\begin{align}
f_h=-\square f_0,\quad a_h=-\square a_0,\quad t_h=-\square t_1.
\end{align}
They do not encode dynamics of the theory. The corresponding logarithmic and finite terms can be removed by the following counter terms with appropriate normalizations
\begin{align}
&\cN\sqrt{-\g}\d\th\square\d\th\ln\L,\quad \cN\sqrt{-\g}a_t\square a_t\ln\L, \quad \cN\sqrt{-\g}\ph\square\ph\ln\L, \no
&\cN\sqrt{-\g}\d\th\square\d\th,\quad \cN\sqrt{-\g}a_t\square a_t, \quad \cN\sqrt{-\g}\ph\square\ph,
\end{align}
Adding all counter terms to \eqref{S_pd} and dropping contributions like $m^2t_1^2$, which do not encode dynamics of the theory, we obtain the following renormalized on-shell action
\begin{align}\label{S_renorm}
S^{\text{ren}}=\cN r_0^4\(-\frac{t_1t_3}{4}+\frac{a_0a_2}{2}-\frac{m^2f_0f_2}{4}\).
\end{align}
Now we can do variation of \eqref{S_renorm} with respect to sources to obtain vev of the corresponding operators. Note that partial derivatives hit twice in each terms in the bracket. We obtain 
\begin{align}
&\d\s=\frac{\d S^{\text{ren}}}{\d t_1}=\cN r_0^3(2\pi\a')\(-\frac{t_3}{2}\), \no
&\d n=\frac{\d S^{\text{ren}}}{\d a_0}=\cN r_0^3(2\pi\a')a_2, \no
&\s_5=\frac{\d S^{\text{ren}}}{\d f_0}=\cN r_0^4\(-\frac{m^2f_2}{2}\), \no
\end{align}
We use the $\d$ symbol to indicate that the vev is on top of a nonvanishing background. Taking the derivatives once more, we obtain correlators shown in \eqref{dict}.

\bibliographystyle{unsrt}
\bibliography{Q5ref}

\end{document}